\documentclass[letterpaper, 10 pt, conference]{ieeeconf}  

\IEEEoverridecommandlockouts                              

\usepackage{cite}
\usepackage[english]{babel}
\usepackage[utf8]{inputenc}
\usepackage[cmex10]{amsmath}
\usepackage{amssymb}
\usepackage[bottom]{footmisc}
\usepackage{float}
\usepackage[pdftex]{graphicx}
\usepackage{amsfonts}
\usepackage{mathtools}
\usepackage{xcolor}
\usepackage{color}
\usepackage{verbatim}
\usepackage{graphicx}
\usepackage{caption}
\usepackage{subcaption}
\usepackage{tcolorbox}
\usepackage{soul}
\usepackage{xcolor}
\usepackage{tcolorbox}
\tcbuselibrary{raster}
\usepackage[norelsize,linesnumbered,boxed,vlined]{algorithm2e}
\graphicspath{{./figures/}}

\usepackage{stlcbf}
\usepackage{setspace}
\usepackage{booktabs}

\let\labelindent\relax
\usepackage{enumitem}

\renewcommand{\baselinestretch}{0.97}
\title{{\Large Risk-Awareness in Learning Neural Controllers for
Temporal Logic Objectives}}

\author{
\authorblockN{Navid Hashemi\authorrefmark{1}, %
              Xin Qin\authorrefmark{1}, %
              Jyotirmoy V. Deshmukh\authorrefmark{1}, \\ %
              Georgios Fainekos\authorrefmark{2}, %
              Bardh Hoxha\authorrefmark{2},
              Danil Prokhorov\authorrefmark{2},
              Tomoya Yamaguchi\authorrefmark{2}}
\authorblockA{\authorrefmark{1}University of Southern California,
              \authorrefmark{2}Toyota Motor North America R\&D.}
}

\begin{document}

\maketitle
\thispagestyle{empty}
\pagestyle{empty}

\begin{abstract}
In this paper, we consider the problem of synthesizing a controller in the presence of uncertainty such that the resulting closed-loop system satisfies certain hard constraints while optimizing certain  (soft) performance objectives. We assume that the hard constraints encoding safety or mission-critical task objectives are expressed using Signal Temporal Logic (STL), while performance is quantified using standard cost
functions on system trajectories. In order to prioritize the satisfaction of the hard STL constraints, we utilize the framework of control barrier functions (CBFs) and algorithmically obtain CBFs for STL objectives. We assume that the controllers are modeled using neural networks (NNs) and provide an optimization algorithm to learn the optimal parameters for the NN controller that optimize the performance at a user-specified robustness margin for the safety specifications. We use the formalism of risk measures to evaluate the risk incurred by the trade-off between robustness margin of the system and its performance.
We demonstrate the efficacy of our approach on well-known difficult examples for nonlinear control such as a quad-rotor and a unicycle, where the mission objectives for each system include hard timing constraints and safety objectives. 


\end{abstract}

\section{Introduction}
\label{sec:intro}
Safety-critical cyber-physical systems typically have hard safety
specifications that must be met by all system behaviors to guarantee
system safety. Additionally, due to efficiency concerns, system
designers often specify performance objectives, and seek controllers
to optimize these objectives. For example, consider an autonomous
vehicle (AV) following another vehicle. Here, the AV must satisfy the
safety specification of maintaining a minimum safe distance
($d_\mathit{safe}$) from the lead vehicle. However, the system
designer may also want to minimize the travel time for the AV.
Clearly, the vehicle can be safe with a high robustness margin by
driving slower than required (maintaining distance much greater than
$d_\mathit{safe}$), but this leads to sub-optimal performance w.r.t.
the travel time objective. In many cases, designing for safety and
performance objectives may require design trade-offs. While 
designers must never violate safety requirements in favor of
performance, they can trade-off the safety margin against performance.
This trade-off thus generates some risk: from a safety perspective how
risky is to use a controller that may perform better with a lower safety margin? 
In this paper, we systematically study this problem. 

We assume that safety specifications are provided in a real-time
temporal logic such as {\em Signal Temporal Logic} (STL)
\cite{maler2004monitoring}.  STL has recently emerged as a powerful
specification language in the various cyber-physical system
applications
\cite{pant_iccps,raman2014model,monitoring_chapter,aksaray_q-learning_2016}.
In STL properties, predicates over real-valued signals form atomic
subformulae which can be combined using Boolean logic connectives
(such as and, or, not), and temporal logic operators (such as
eventually, always, until) that are indexed by time intervals.  For
example consider a design objective for a quadcopter: ``The quadcopter
must rendezvous in one of two designated regions $R_1$ or $R_2$
exactly 5 to 7 minutes after takeoff before getting as close as
possible to a given target destination within 20 mins, while avoiding
no-fly zones.'' Let $p(\cdot)$ denote the position of the quadcopter.
The hard safety specifications in this objective can be expressed by
the following STL formula:
\begin{equation}\label{stl:qc}
    \varphi_{qc}\ \equiv\ \ev_{[5,7]}(p\in R_1  \vee p \in R_2)  \wedge \alw_{[0,20]} (p
    \not\in R_{\mathit{nofly}})
\end{equation}
The soft specification requires us to minimize
$d(p,p_{\mathit{target}})$, where $d$ is a distance function and
$p_\mathit{target}$ is the target. An advantage of STL is that we can
quantify how robustly a given system behavior satisfies an STL
property using the notion of a {\em robustness} value
\cite{FainekosP06fates}. Given a system behavior and a specification,
the robustness can be thought of as a signed distance from the given
system behavior to the set of behaviors satisfying the property. We
can say that a system has safety robustness margin $\rho^* > 0$ if the
minimum robustness value across all its system behaviors exceeds
$\rho^*$. We assume that a performance objective is specified as any
differentiable, real-valued function of the system behavior. 

There has been considerable amount of research on the problem of
synthesizing controllers that guarantee STL specifications.  For
example, using approaches from motion planning
\cite{karaman2011sampling, shoukry2016scalable}, model predictive
control \cite{raman2014model,farahani2015robust,gol2015temporal},
reactive synthesis \cite{raman2015reactive, lindemann2021reactive },
reinforcement learning
\cite{berducci2021hierarchical,li2017reinforcement}, imitation
learning \cite{liu2021recurrent,YaghoubiF2019tecs}, and through the
use of {\em control barrier functions} \cite{lindemann2018control,
lindemann2019robust}.  Of these approaches, the most relevant to our
paper is the one based on using control barrier functions (CBFs)
\cite{xu2015robustness}. A CBF describes a set $C$ such that for all
system states $\state_\timeid \in C$, there exists a control action
that ensures that $\state_{\timeid+1} \in C$.  Control synthesis from
CBFs has seen a lot of recent work
\cite{ames2019control,nilsson2018barrier,xu2015robustness}.  Recent
work has focused on CBFs that provide more general classes of
invariants such as timed reachability \cite{garg_acc2020,garg_cdc19}
and fragments of STL \cite{stl_cbf}.  
{\em Prima facie}, synthesis of controllers to satisfy STL specifications may
look like a well-studied problem, however, several open problems remain:
\begin{enumerate}[nosep,wide,labelwidth=!]
\item
Existing work may use hand-crafted CBFs over limited fragments of STL
\cite{lindemann2018control}. E.g., existing work does not address
disjunctive STL specifications (see Eq. \eqref{stl:qc}). 

\pagebreak[2]

\item
Existing approaches do not consider the trade-off between safety and
performance. A na\"{i}ve encoding of the problem using Lagrange
multipliers (as we show in this paper) does not scale, thus
demonstrating the need for a more nuanced approach.
\pagebreak[2]

\item
Many existing approaches focus on control of linear systems or simple
nonlinear systems.
\pagebreak[2]

\item
Existing work does not quantify risk awareness in trading off safety
margin versus system performance.
\end{enumerate}

To address all the above challenges, we first formulate an objective
function that combines a CBF for STL-based safety specifications with
performance objectives using Lagrange multipliers.  We then
demonstrate that the Lagrangian optimization approach does not scale.
We provide an algorithm to automatically generate the CBF for the STL
specification directly from its structure. An important consideration
in the CBF is our use of the {\em weighted average} of subformula CBFs
to generate the CBF for a disjunctive formula. 

Next, we introduce deep neural network (DNN)-based controllers to
handle arbitrary nonlinear systems. We train the DNN-based controllers
in a model-free fashion using a stochastic gradient optimization method
that uses adaptive moments.  Our optimization formulation is similar
to the problem of training a recurrent neural network (RNN), where a
cascade of NNs for a given temporal horizon is trained.  A crucial
aspect of our optimization algorithm is to explicitly guide the search
for DNN parameters using a robustness margin parameter: across
iterations, the optimizer alternates between satisfying safety and
performance based on the robustness of the DNN controller
vis-\`{a}-vis the desired robustness margin.  


Finally, we evaluate the risk-awareness for each designed controller
by picking different robustness margins as design parameters. For this
analysis, we utilize the recently formulated risk-aware verification
approach \cite{akella2022scenario} that uses risk measures such as
value-at-risk and conditional-value-at-risk. We demonstrate the
efficacy of our method on several examples of nonlinear
systems and disjunctive STL safety specifications.

\section{Background}
\label{sec:background}
In this section, we provide the mathematical notation and the overall
problem definition.  We use bold letters to indicate vectors and
vector-valued functions, and calligraphic letters to denote sets.

Let $\state$ and $\action$ respectively be the variables denoting
state and control inputs taking values from compact sets $\states
\subseteq \reals^n$ and $\action \subseteq \reals^m$, respectively. We
use the words {\em action} and control input interchangeably. We
consider discrete-time nonlinear feedback control systems of the
following form\footnote{Our technique can handle continuous-time
nonlinear systems as well. This requires zero-order hold
discretization of the dynamics in a sound way to account for system
behavior between sample times.}:
\begin{equation}
\label{eq:discdynamics}
    \state_{\timeid+1} = \dynamics(\state_\timeid,\action_\timeid).
\end{equation}
Here, $\state_\timeid$ and $\action_\timeid$ denote the values of the
state and action variables at time $\timeid$.
We assume that the controller can be expressed as a parameterized function 
$\policy_\param$, where $\param$ is a vector of parameters that takes
values in $\params$.  Later in the paper, we instantiate the specific
parametric form using a neural network for the controller.  Given a
fixed vector of parameters $\param$, the parametric control policy
$\policy_\param$ returns an action $\action_\timeid$ as a function of
the current state $\state_\timeid \in \states$ and time $\timeid \in
\posint$.  Namely, 
%
%
%
\begin{equation}
\label{eq:DNNctrl}
\action_\timeid = \policy_\param(\state_\timeid, \timeid)
\end{equation}
We will be using the terms controller and control policy interchangeably.
Under a fixed policy, Eq.~\eqref{eq:discdynamics} is an autonomous
discrete-time dynamical system. For a given initial state $\state_0
\in \init \subseteq \states$ and dynamics $\dynamics$, a system
trajectory $\traj^{\param}_{\state_0,\dynamics}$ is a function from
$[0,\horizon] \subset \pposint$ to $\states$, where
$\traj^{\param}_{\state_0, \dynamics}(0) = \state_0$, and for all
$\timeid \in [0,\horizon-1]$, $\traj^{\param}_{\state_0,
\dynamics}(\timeid+1) =
\dynamics(\state_\timeid,\pi_\param(\state_\timeid, \timeid))$. To
address modeling inaccuracies, we also consider bounded uncertainty in
the model.  We denote $\modelfamily$ as the family of possible
realizations of the model $\dynamics \in \modelfamily$. 
If the policy $\policy_\param$ is obvious from the context, we drop
the $\param$ in the notation $\traj^{\param}_{\state_0,\dynamics}$.
The main objective of this paper is to formulate algorithms to obtain
the optimal policy $\optpolicy$ that guarantees the satisfaction of
certain task objectives and safety constraints while optimizing
performance rewards. In the rest of the section, we formulate
controller synthesis as an optimization problem that we seek to solve.
In order to define this formally, we first introduce a performance
reward, and then introduce task objectives/safety constraints. 

\mypara{Performance reward}
In practical control applications, it is common to quantify the control performance
using a state-based reward function 
\cite{anderson2007optimal},\cite{sutton2018reinforcement}. Formally,
\begin{definition}[Performance reward for a trajectory]
Given a reward function $\statecost : \states \to \reals$, and discount
factor, $\gamma \in [0,\ 1]$, the performance reward of a trajectory
initiating in state $\state_0$, under policy $\pi_\param$ is defined
in Eq.~\eqref{eq:perfcost}.
\begin{equation}
\label{eq:perfcost}
\perfobjectiveTraj{\state_0,\dynamics} = 
    \sum_{\timeid=0}^{\horizon} \gamma^{\timeid}
        \statecost(\traj^{\param}_{\state_0,\dynamics}(\timeid))
\end{equation}
\end{definition}



\mypara{Task Objectives and Safety Constraints}
We assume that task objectives or safety constraints of
the system are specified in a temporal logic known as 
Signal Temporal Logic (STL)\cite{maler2004monitoring}. STL 
formulas are defined using the following syntax:

\begin{equation}
\label{eq:stlfrag}
\resizebox{\hsize}{!}{$
\varphi\  =\  h(\state) \bowtie 0 \mid 
             \varphi_1 \wedge \varphi_2 \mid
             \varphi_1 \vee \varphi_2 \mid
             \ev_\intvl \varphi \mid
             \alw_\intvl \varphi \mid
             \varphi_1 \until_\intvl \varphi_2
$}
\end{equation}
Here, $\bowtie \in \{ \le, <, >, \ge \}$, $h$ is a function from
$\states$ to $\reals$, and $\intvl$ is a closed
interval $[a,b] \subseteq [0,\horizon]$.  



\myipara{Semantics} The formal semantics of STL over discrete-time
trajectories have been previously discussed in
\cite{FainekosP06fates}. 
We denote the formula $\varphi$ being true at
time $\timeid$ in trajectory $\traj_{\state_0,\dynamics}$ by
$\traj_{\state_0,\dynamics},\timeid \models \phi$. We say that
$\traj_{\state_0,\dynamics},\timeid \models h(\state) \bowtie 0$ iff
$h(\traj_{\state_0,\dynamics}(\timeid)) \bowtie 0$. The semantics of
the Boolean operations ($\wedge$, $\vee$) follow standard logical
semantics of conjunctions and disjunctions respectively. 
For temporal operators, we say $\traj_{\state_0,\dynamics} \models
\ev_\intvl \varphi$ is true if there is a time $\timeid \in \intvl$
where $\varphi$ is true.  Similarly, $\traj_{\state_0,\dynamics}
\models \alw_\intvl \varphi$ is true iff $\varphi$ is true for all
$\timeid \in \intvl$. Finally, $\traj_{\state_0,\dynamics} \models
\varphi_1 \until_\intvl \varphi_2$ if there is a time $\timeid \in
\intvl$ where $\varphi_2$ is true and for all times $\timeid' \in
[0,\timeid)$ $\varphi_1$ is true. 

In addition to the Boolean satisfaction semantics, STL also permits
quantitative satisfaction semantics. These are defined with a
{\em robustness} function $\rob$ evaluated over a trajectory. We omit the formal
definition; it can be found in
\cite{FainekosP06fates,maler2004monitoring}. Intuitively, the
robustness function defines robustness of predicates at a given time
$\timeid$ to be proportional to the signed distance of the state
variable value at $\timeid$ from the set of values satisfying the
predicate. Conjunctions and disjunctions map to minima and maxima of
the robustness of their subformulas respectively. Temporal operators
can be viewed as conjunctions/disjunctions (or their combinations) over
time. We denote $\rob_\varphi(\state_0,\dynamics)$ as the robustness
of the trajectory starting in state $\state_0$ for the dynamics
$\dynamics$. Note that if $\rob_\varphi(\state_0,\dynamics) > 0$ then
it implies that $\traj_{\state_0,\dynamics} \models \varphi$.

\mypara{Risk Measures} We now introduce two commonly used risk
measures that are used to provide probabilistic guarantees of system
correctness. We assume that we are provided with a probability
distribution $\distribution_\modelfamily$ on model uncertainties (which is a
distribution on $\modelfamily$), and $\distribution_\init$ to 
denote a distribution over the initial states of the system. 
A risk measure 
at a given threshold $\confthreshold$ (denoted
$\riskm_\confthreshold$) is a quantity that can be used to provide
the following probabilistic guarantee about the robustness of a
given STL specification for the system:
\begin{equation}
\label{eq:probrisk}
\state_0 \sim \distribution_\init, \dynamics \sim
\distribution_\modelfamily \implies \Pr(
-\rob_\varphi(\state_0,\dynamics) \le \riskm_\confthreshold) \ge
\confthreshold
\end{equation}



We now include two of the standard risk measures used in
literature from \cite{risk_measure}. 

\begin{definition}[Value-at-Risk ($\vaar$), Conditional-Value-at-Risk ($\cvar$) \cite{risk_measure}]
Let $Z$ be shorthand for $\rob_\varphi(\state_0,\dynamics)$. The
Value-at-Risk is defined as follows: 
\begin{equation}
    \vaar_\confthreshold(-Z) =
     \inf_{\zeta \in \reals} \setof{ \zeta \mid \Pr(-Z \le \zeta) \ge \confthreshold }
\label{eq:psi_x_zeta}
\end{equation}
The conditional-value-at-risk is defined as follows:
\begin{equation}
    \cvar_\confthreshold(-Z) = \expectedvalueover{-Z \ge
    \vaar_\confthreshold(-Z)}{-Z}
\end{equation}
\label{def:cvar}
\end{definition}
Essentially, both risk measures provide probabilistic upper bounds on 
the negative of the robustness value, or provide lower bounds on the 
actual robustness value, as is required in risk-aware verification 
\cite{akella2022scenario,risk_measure}.



\mypara{Problem Definition}
(i) Learn an optimal policy
$\policy_{\param}(\state_\timeid,\timeid)$ such that it satisfies a given
STL formula $\varphi$ while maximizing the performance reward defined in Eq.~\eqref{eq:perfcost}.
\begin{equation}
\label{eq:probdef}
\begin{array}{llll}
\optparam = & \! 
\displaystyle\arg\max_{\param \in \params} & \quad & \expectedvalueover{\state_0,\dynamics \sim \uniformdist}{\perfobjectiveTraj{\state_0,\dynamics}} \\  
& \! \text{s.t.}        & \quad & \forall \state_0 \in \init,\ \forall \dynamics \in \modelfamily: \traj_{\state_0,\dynamics} \models \varphi
\end{array}
\end{equation}

(ii) Given a confidence threshold $\confthreshold$, we will compute the risk measure $\riskm_\confthreshold$ that guarantees that:
$ (\state_0, \dynamics) \sim (\distribution_\init \times \distribution_{\modelfamily}) 
\implies \Pr( -\rob_\varphi(\state_0,\dynamics) \le
\riskm_\confthreshold ) \ge \confthreshold$.

\section{Control Barrier Functions for STL}
\newcommand{\barrier}{b}
\newcommand{\barrierof}[1]{\barrier_{#1}}
\newcommand{\zerolevelset}{B}
\newcommand{\dmax}{\displaystyle\max}
\newcommand{\softmin}{\mathsf{softmin}}
\newcommand{\selector}{\mathsf{sel}}
\newcommand{\ignore}[1]{}

In \cite{lindemann2018control}, the authors introduce {\em
time-varying control barrier functions} that are used to synthesize
controllers that are guaranteed to satisfy a given STL specification.  We first adapt this notion to discrete-time
nonlinear systems. 
\begin{definition}[Discrete-Time Time-Varying Valid Control Barrier Functions (DT-CBF)]
Let $\barrier : \states \times [0,\horizon] \rightarrow \reals$ be a
function that maps a state and a time instant to a real value.  Let
$\zerolevelset(\timeid) = \{\state_\timeid \mid
\barrier(\state_\timeid,\timeid) \ge 0 \}$ be a time-varying set.  The
function $\barrier$ is a valid, discrete-time, time-varying CBF if it
satisfies the following condition: \\

The zero levelsets of the CBF are an envelope for any
system trajectory, i.e.,
\begin{equation}
\label{eq:barrier_env}
\forall \state_0 \in \init,\ \forall \dynamics \in \modelfamily :\forall \timeid \in [0,\horizon]: \traj_{\state_0,\dynamics}(\timeid) \in \zerolevelset(\timeid)
\end{equation} 

\end{definition}

\SetKwProg{Fn}{Function}{}{end}
\newcommand{\mysty}[1]{$\mathtt{#1}$}
\SetKwSwitch{Switch}{Case}{Other}{switch}{}{case}{otherwise}{endcase}{endsw}
\SetFuncSty{mysty}
\SetKw{Return}{return}
\DontPrintSemicolon
\begin{algorithm}[t]
\SetKwFunction{stltocbf}{stl2cbf}
\Fn{\stltocbf{$\varphi$,$\traj_{\state_0,\dynamics}$}}{
    \uCase{$\varphi = h(\state,\timeid) \ge 0$}{
       \Return $\predbarrier(h(\state_\timeid,\timeid))$ \;
    }
    \uCase{$\varphi=\varphi_1 \wedge \varphi_2$}{
       \Return $\resizebox{1.04\hsize}{!}{$\softmin\left(\stltocbf{$\varphi_1$,$\traj_{\state_0,\dynamics}$}, \stltocbf{$\varphi_2$,$\traj_{\state_0,\dynamics}$};\eta\right)$}$\;
    }
    \uCase{$\varphi=\varphi_1 \vee \varphi_2$}{
       \Return $\resizebox{1.04\hsize}{!}{$\weightedaverage\left(\stltocbf{$\varphi_1$,$\traj_{\state_0,\dynamics}$}, \stltocbf{$\varphi_2$,$\traj_{\state_0,\dynamics}$};\bbeta^\varphi\right)$}$\;
    }
    \uCase{$\varphi=\alw_{[a,b]}\varphi$}{
       \Return $\underset{\timeid \in [a,b]}{\softmin}\left(\stltocbf{$\varphi$,$\traj_{\state_0,\dynamics}$};\eta \right)$ \;
    }
    \uCase{$\varphi=\ev_{[a,b]}\varphi$}{
       \Return $\underset{\timeid\in[a,b]}{\weightedaverage}(\stltocbf{$\varphi$,$\traj_{\state_0,\dynamics}$};\bbeta^\varphi)$ \;
    }
    \uCase{$\varphi=\varphi_1\unt_{[a,b]}\varphi_2$}{
        \For{$k \gets a$ to $b$} {
                $$
                \resizebox{1\hsize}{!}{$
                \begin{aligned}
                &\stltocbf{$\varphi,\traj_{\state_0,\dynamics}$} \gets \weightedaverage\Big(\stltocbf{$\varphi,\traj_{\state_0,\dynamics}$},\\
                &\softmin\big(\stltocbf{$\varphi_2,\traj_{\state_0,\dynamics}$},\\
                &\stltocbf{$\alw_{[0,\timeid-1]}\varphi_1, \traj_{\state_0,\dynamics}$}; \eta \big); \bbeta^\varphi\Big)
                \end{aligned}
                $}
                $$
        } \;
        \Return \stltocbf{$\varphi,\traj_{\state_0,\dynamics}$}\;
    }
}
\caption{\small{Recursive formulation of CBFs based on an STL formula}}
\label{algo:stl2cbf}
\end{algorithm}
\myipara{DT-CBF for STL}
We formulate CBFs in a recursive fashion based on the formula structure. We describe the
overall procedure in Algorithm~\ref{algo:stl2cbf}.
Before we describe the actual algorithm, we introduce some helper functions. 
The $\softmin$ function
defined in Eq.~\eqref{eq:smoothmin} has been used in the past by several approaches \cite{YaghoubiF2019tecs,pant2017smooth,leung2019backpropagation} 
as a smooth approximation for computing the minimum of a number of real-valued quantities. 
In our $\softmin$ function, we introduce an additional parameter $\eta > 1$ that is used to control
the level of conservatism in the approximation. Intuitively, as larger values of $\eta$ reduce the 
conservatism but require greater numeric precision. Later, we discuss how the $\softmin$ function
appears as a part of a cost function to be optimized; we include $\eta$ as a part of this optimization
process.

\begin{equation}\label{eq:smoothmin}
\softmin(v_1,\ldots,v_k;\eta) = -\frac{1}{\eta} \ln\left(
                                \sum_{i=1}^{k} e^{-\eta v_i}
                                             \right)
\end{equation}

We also define the weighted average function $\weightedaverage$. We are interested in two different form of $\weightedaverage$ presented in Eq.~\eqref{eq:wtavg1} and Eq.~\eqref{eq:wtavg2}; here $\bbeta = (\beta_1,\ldots,\beta_k)$.
\begin{equation}
\label{eq:wtavg1}
\weightedaverage_1(v_1,\ldots,v_k;\bbeta) = \sum_{i=1}^k \left(\frac{\beta_i^2}{\sum_{i=1}^k \beta_i^2}\right) v_i,
\end{equation}
\begin{equation}
\label{eq:wtavg2}
\weightedaverage_2(v_1,\ldots,v_k;\bbeta) =\sum_{i=1}^k\left(\frac{ \exp(\beta_i) }{\sum_{j=1}^k \exp(\beta_j)}\right)v_i,
\end{equation}

\noindent where the former is more accurate but the latter is more efficient for gradient descent. Finally, we articulate useful properties of $\softmin$ and $\weightedaverage$ in Lemma~\ref{lem:softminwtavg}.

\begin{lemma}
\label{lem:softminwtavg}
For all $v_1,\ldots,v_k \in \reals$, and for $\eta \in \reals_{>1}$, 
the following are true:
\begin{eqnarray}
& (\min_i v_i) \ge \softmin(v_1,\ldots,v_k;\eta) \label{eq:softminlemma} \\
& 
        (\min_i v_i) \le \weightedaverage(v_1,\ldots,v_k;\bbeta) \le (\max_i v_i)
 \label{eq:wtavglemma}
\end{eqnarray}
\end{lemma}

We can now describe Algorithm~\ref{algo:stl2cbf}. The function
$\barrierof{\phi}$ computes the CBF w.r.t. either an atomic signal
predicate or a conjunction of atomic predicates. The CBF for an atomic
predicate $\phi$ of the form $h(\state_\timeid)>0$ is defined using a
function $\predbarrier$ that ensures that
$\predbarrier(h(\state_\timeid))$ is positive if $h(\state_\timeid)$
is positive, $0$ if it is zero and negative otherwise.  The CBF of the
conjunction of two predicates is simply the $\softmin$ of the CBFs of
the conjuncts. In the function
$\stltocbf(\varphi,\traj_{\state_0,\dynamics})$, we consider four
cases.  If $\varphi$ is a formula of the form $\alw_{[a,b]}\phi$, then
we return the $\softmin$ of $\barrierof{\phi}(\state_\ell,\ell)$ for
all $\ell \in [a,b]$. If $\varphi$ is of the form $\ev_{[a,b]}\phi$,
then we return the weighted average of the CBFs at all time instants
in $[a,b]$. For conjunctions of either kinds of temporal formulas, we
again return the $\softmin$ and for disjunctions, we return the
weighted average. Note that the function $\stltocbf$ can be invoked
with a concrete trajectory whereupon it  returns a numeric value. It
can also be invoked with a symbolic trajectory (where the symbols
$\state_\timeid$ indicate the symbolic state at time $\timeid$),
whereupon it returns a symbolic candidate CBF that is the smooth
robustness of the trajectory and is a guaranteed lower bound for
trajectory robustness $\rob_\varphi(\state_0, \dynamics)$.
 
\begin{lemma}
For any formula $\varphi$ belonging to STL, for a given 
trajectory $\traj_{\state_0,\dynamics} = \state_0,\state_1,\ldots,\state_n$,
if $\stltocbf(\varphi,\traj_{\state_0,\dynamics}) > 0$, then $\traj_{\state_0,\dynamics} \models \varphi$.
\end{lemma}

\begin{proof}
We can prove this recursively over the formula structure and from the identities in Lemma~\eqref{lem:softminwtavg}. \textcolor{black}{ It is necessary to mention if $\stltocbf()<0$ it does not imply the STL specifications are violated.}
\end{proof}

\newcommand{\region}{\mathcal{E}}

\begin{example}
\label{ex:one}
Consider the STL specification in Eq.~\eqref{eq:stlspec}.
\begin{equation}
\label{eq:stlspec}
\left(\ev_{[1,10]}\left( \state \in \region_1 \right) \vee 
      \ev_{[1,10]} \left( \state \in \region_2 \right) \right)   \wedge 
\alw_{[1,20]} \left( \state \not\in \region_3 \right)
\end{equation}
Let $c_2 = (2,8)$, $c_1 = (5,5)$, $c_3 = (8,2)$ and $r=\sqrt{1.5}$. Then,
in Eq.~\eqref{eq:stlspec}, for $i \in [1,3]$: $\region_i = 
\transpose{(\state-c_i)}(\state-c_i) \le r$. 
We define the 
CBF $\predbarrier(\state_\timeid \in \region_1)$ to be 
\mbox{$(1-e^{-(\transpose{(\state_\timeid-c_1)}(\state_\timeid-c_1)-r)})$}. 
For $j = 2,3$, we define $\predbarrier(\state_\timeid \in \region_j)$ to be 
$r-\transpose{(\state_\timeid - c_j)}(\state_\timeid - c_j)$. Then, the 
CBF w.r.t. the formula can be computed using Algorithm.~\ref{algo:stl2cbf}.
\end{example}

\section{Learning-based Control Synthesis}
\label{sec:learning}

We remark that the trajectory
$\traj_{\state_0,\dynamics}$ is essentially a repeated composition of
$\dynamics$ and the neural controller. Thus, we can compute the
gradient of the performance costs and STL objectives (as expressed by
the CBF) with respect to the controller parameters $\param$ using
standard backpropagation methods for neural networks.

\subsection{Training Neural Networks to satisfy specifications}

We explain the procedure for training a neural controller w.r.t. performance and safety specifications in algorithm~\ref{algo:training}. The training algorithm aims to approximate the solution of Eq.~\eqref{eq:probdef}. Thus, the first step is to reformulate it free from constraints. Algorithm \ref{algo:stl2cbf} provides the smooth trajectory robustness for a given STL formula $\varphi$. This robustness is a function of the common variable $\eta$ that is used by all $\softmin$ functions, and the tuple of $\beta$ variables for each subformula $\psi$ of a disjunctive formula (denoted  $\bbeta^\psi$). In addition to the neural network parameters $\param$, we also treat the variables $\eta$ and the sets $\{\bbeta^\psi\}$ as decision variables in the training process. The $\beta$ variables are already unconstrained; however, $\eta$ is constrained: $\eta > 1$. To remove this constraint, we introduce $\eta = \lambda^2 + 1$, and use $\lambda$ as a decision variable. We denote the tuple of conjunctive variable ($\lambda$) and all disjunctive variables ($\bbeta^\psi$) with $\vars$. We also denote this robustness with $\STLobjective$ that is a function of tuple $(\state_0,\  \dynamics, \ \param,\ \vars)$,
\begin{equation}\label{eq:CBF}
\STLobjective(\state_0,\dynamics,\param,\vars) = \stltocbf(\varphi, \traj_{\state_0,\dynamics}),\ \
\vars = (\{\bbeta^\psi\},\lambda).
\end{equation}
We also sample a batch  $\sampledinits$ of initial states uniformly from $\init$ for training purposes. 

The training algorithm is primarily inspired by the Lagrange multiplier technique that transforms a constrained optimization to non-constrained,
\begin{equation}\label{eq:lagrangian}
\objective = \underset{\param, \vars}{\max}
       \sum_{\state_0 \in \sampledinits, \dynamics \in \modelfamily } \left(
             \perfobjective(\state_0,\param) +
             \omega_{\state_0}\STLobjective(\state_0,\param,\vars)\right).
\end{equation} 
First order optimality conditions guarantee that as long as the Lagrange multipliers are positive, the cost as defined using $\STLobjective$ is positive. This in turn guarantees the satisfaction of STL specifications along the trajectory. Since $\STLobjective(\state_0,\dynamics,\param,\vars)$ is highly non-convex, optimization \eqref{eq:lagrangian} is quite intractable and the solution may not satisfy the KKT optimality condition \cite{boyd2004convex}. However, the main role of the Lagrange multipliers $\omega_{\state_0,\dynamics}$ is to perform a trade-off between $\STLobjective$ and $\perfobjective$ and one of the contributions of this work is to propose a training process that focuses on applying this trade off. The training algorithm utilizes the gradients $\nabla_\param \perfobjective(\state_0,\dynamics,\param)$ and $\nabla_{\param, \vars} \STLobjective(\state_0,\dynamics,\param, \vars)$ (obtainable from Automatic differentiation package \cite{paszke2017automatic}). In case, the cost specified with $\STLobjective$ is less than a user-specified threshold, then the algorithm increases this with a wise selection between $\nabla_\param \perfobjective(\state_0,\dynamics,\param)$ and $\nabla_{\param, \vars} \STLobjective(\state_0,\dynamics,\param, \vars)$ . Otherwise, it increases the performance with $\nabla_\param \perfobjective(\state_0,\dynamics,\param)$. We call the mentioned user specified threshold as {\em robustness margin} (denoted by $\rho$).    
\RestyleAlgo{ruled}
\begin{algorithm}[t]
$i \gets 0$, \textbf{Initialize} $(\param, \vars)^0$, \textbf{Sample} $\init$ to obtain $\sampledinits$ \;
\While{true}{
    \textbf{Sample} $\dynamics \in \modelfamily$ \nllabel{algoline:samplemodel}\;  
    \ForEach{$\state_0 \in \sampledinits$}{
        \nllabel{algoline:startcomputegrads}
        $\delta_1(\state_0) \gets
        [\gradient{\perfobjective(\state_0,\dynamics,\param)}{\param},\ 
        0]$ \;
        $\delta_2(\state_0) \gets
        [\gradient{\STLobjective(\state_0,\dynamics,\param,\vars)}{\param},\ 
        \gradient{\STLobjective(\state_0,\dynamics,\param,\vars)}{\vars}]$ \;
    } \nllabel{algoline:endcomputegrads}
    
    \tcp{$\mathtt{get\ states\ with\ the\ best\ grad.\ values}$}
    $b1, b2 \gets
        \displaystyle\argmax_{\state_0\in\sampledinits}
            {\twonorm{\delta_1(\state_0)}}$, 
        $\displaystyle\argmax_{\state_0\in\sampledinits}
            {\twonorm{\delta_2(\state_0)}}$ \;
    $d_1,d_2 \gets \delta_1(\textrm{b1}), \delta_2(\textrm{b2})$ \nllabel{algoline:d1d2} \;
    \tcp{$\mathtt{candidate\ parameter\ updates}$}
    \setstretch{1.1}
    $(\param_{1},\vars_{1}) \gets
        (\param,\vars)^i+\adam(d_1)$ \nllabel{algoline:perfupdate} \;
    $(\param_{STL},\vars_{STL}) \gets
        (\param,\vars)^i+\adam(d_2)$ \nllabel{algoline:stlupdate} \;
    $(\param_{1,slow},\vars_{1,slow}) \gets
        (\param,\vars)^i+\adam(d_1)/\tau$ \nllabel{algoline:penalty} \;
   \textbf{Sample} $\state_0^i$ from $\init$\;
   \tcc{$\mathtt{Pick\ update\ giving\ best\ tradeoff}$ $\mathtt{\ between\
   perf.\ and\ safety}$}
   \If{$\STLobjective(\state_0^i,\dynamics,(\param,\vars)^i) \le \rho $}{ \nllabel{algoline:firstif}
        \If{$\STLobjective(\state_0^i,\dynamics,\param_{1},\vars_{1}) \ge 
             \STLobjective(\state_0^i,\dynamics,(\param,\vars)^i)$}
            {
                $(\param,\vars)^{i+1} \gets 
                 (\param_{1}, \vars_{1})$ \nllabel{algoline:bothimproved} \; 
        } \Else {
                $(\param,\vars)^{i+1} \gets
                 (\param_{STL},\vars_{STL})$ \nllabel{algoline:stlimproved} \;
        }
   } \Else {
                $(\param,\vars)^{i+1} \gets 
                 (\param_{1,slow}, \vars_{1,slow})$ \nllabel{algoline:cautious} \;
    
   }
}
\caption{Sampling based algorithm for training the parameterized policy.}
\label{algo:training}
\end{algorithm}

We now describe Algorithm~\ref{algo:training}. We use the variable $i$ to
denote the iteration number during training. We use the notation $(\param,\vars)^i$
to denote the value of $\param$ and $\vars$ at the beginning of iteration $i$.
We initialize $(\param,\vars)^0$ randomly.

At the beginning of each training iteration, in line ~\ref{algoline:samplemodel} we sample $\dynamics$ from $\modelfamily$, then in lines (\ref{algoline:startcomputegrads}-\ref{algoline:endcomputegrads}), for all states in $\sampledinits$, we calculate the gradients 
$\gradient{\perfobjective(\state_0, \dynamics, \param)}{\param}$,
$\gradient{\perfobjective(\state_0,\dynamics, \param)}{\vars}$ (stored in $\delta_1(\state_0)$), 
and  $\gradient{\STLobjective(\state_0,\dynamics, \param,\vars)}{\param}$
$\gradient{\STLobjective(\state_0,\dynamics, \param,\vars)}{\vars}$ (stored in $\delta_2(\state_0)$). 
Of these, note that  $\gradient{\perfobjective(\state_0,\dynamics, \param)}{\vars}$ is $0$.
We then compute the state b1 (resp. b2) for which the $2$-norm of 
$\delta_1(\state_0)$ (resp. $\delta_2(\state_0)$) is the highest. The
gradient values of the states b1 and b2 are respectively
stored in $d_1$ and $d_2$ (Line~\ref{algoline:d1d2}).

The next step is to compute potential updates to the parameter values
$\param$ and the STL parameters $\vars$
(Lines~\ref{algoline:perfupdate}-\ref{algoline:penalty}). Roughly, the
values $(\param_1,\vars_1)$ represent the update to $(\param,\vars)^i$
using only the inclusion of gradient for performance cost in Adam optimizer. The values $(\param_{STL},\vars_{STL})$ represent the
update only using the gradient of smooth trajectory robustness in Adam optimizer. Finally, $(\param_{1,slow},\vars_{1,slow})$ represents a slower update with gradient of performance for some $\tau>1$. 

Next, we sample a state $\state_0^i$ uniformly at random and use it for cost computation.
If $\STLobjective(\state_{0}^i,\dynamics,(\param,\vars)^i) < \rho$, i.e., our
user-provided robustness margin (Line~\ref{algoline:firstif}), then we need to 
take steps to increase the smooth trajectory robustness. We consider two cases: 
\noindent 
(1) If using the update based on the gradient of the performance cost {\em improves} the smooth trajectory robustness, we choose this update as it allows us to improve both performance and robustness, i.e., satisfaction robustness (Line~\ref{algoline:bothimproved}). (2) Otherwise, we use the update based on the gradient of the smooth trajectory robustness $\STLobjective$ (Line~\ref{algoline:stlimproved}).

If $\STLobjective(\state_{0}^i,\dynamics,(\param,\vars)^i) \ge \rho$, then we are robustly satisfying our STL constraints. In further quest to improve the performance cost, we need to take care that we do not reduce the robustness margin w.r.t. STL constraints. Hence, we use a slower learning rate that takes smaller steps in trying to improve the performance (Line~\ref{algoline:cautious}).

\begin{remark}
Considering that this algorithm only focuses on increasing $\STLobjective$ up to $\rho \geq 0$, once the STL specification is satisfied then it focuses on optimizing performance. In a sense, this switching strategy plays a role similar to that of Lagrange multipliers: performance cost is optimized only if the robustness is above the user-provided threshold.
\end{remark}

\subsection{Risk estimation}
The minimum number of samples to guarantee the confidence on the verification results is proposed in \cite{campi2008exact}. We generate $N=10^6$ samples $(\state_0,\dynamics)$ uniformly from $(\init \times \modelfamily)$ and simulate the corresponding trajectories $\traj_{\state_0, \dynamics}$. We compute the robustness $\rob_\varphi(\state_0,\dynamics)$ for every single trajectory and calculate $\vaar$ through obtaining the $\confthreshold*100$ percentile of the negation of the robustness values \cite{calriskusingpercentile} and calculate $\cvar$ according to definition \ref{def:cvar}.

\section{Experimental Evaluation}
\label{sec:experiments}
\subsection{Unicycle Dynamics}
We demonstrate the efficacy of our technique on a nonlinear unicycle model.
We define the uncertainty for the initial condition as:
$$
\resizebox{\hsize}{!}{$
\init=\left\{\state_0 \ | \ (x_0,y_0, \alpha_0)\in \left[ 0.6,\ 1.4 \right] \times \left[0.6,\ 1.4 \right] \times  \left[\frac{2\pi}{5},\ \frac{3\pi}{5} \right] \right\} $}
$$
The unicycle dynamics with uncertainties are defined as follows,

\begin{equation}
\label{eq:beltaexample}
    \begin{bmatrix}x_{\timeid+1}\\ y_{\timeid+1}\\ \alpha_{\timeid+1} \end{bmatrix}=\begin{bmatrix}(1+\delta)x_\timeid+v_\timeid/\omega_\timeid\left( \sin(\alpha_\timeid+\omega_\timeid)-\sin(\alpha_\timeid)\right)\\
    (1+\delta)y_\timeid+v_\timeid/\omega_\timeid\left( \cos(\alpha_\timeid)-\cos(\alpha_\timeid+\omega_\timeid)\right)\\
    (1+\delta)\alpha_\timeid+\omega_\timeid \end{bmatrix},
\end{equation}
\noindent where $\delta \in \left[-0.01,0.01 \right]$ and the control inputs, $v_\timeid,\ \omega_\timeid$ are bounded: $v_\timeid\in[0,\ 1]$, $\omega_\timeid\in[-0.5,\ 0.5]$. To restrict the controller in proposed bounds we fix the last hidden layer of neural controller $[\mathrm{sigmoid}, \tanh]$ and include it to model. Thus, we reformulate the dynamics by replacing the controllers with:
$$
\begin{aligned}
 &v_\timeid &\gets\   &\mathrm{sigmoid} (0.5 a_1(k)), \ &a_1(k)\in \mathbb{R}\\
 &\omega_\timeid &\gets\   &0.5 \tanh(0.5 a_2(k)), \ &a_2(k)\in \mathbb{R}
\end{aligned}
$$


\begin{figure*}
    \begin{minipage}{0.53\textwidth}
    \centering
    \includegraphics[width=0.95\linewidth]{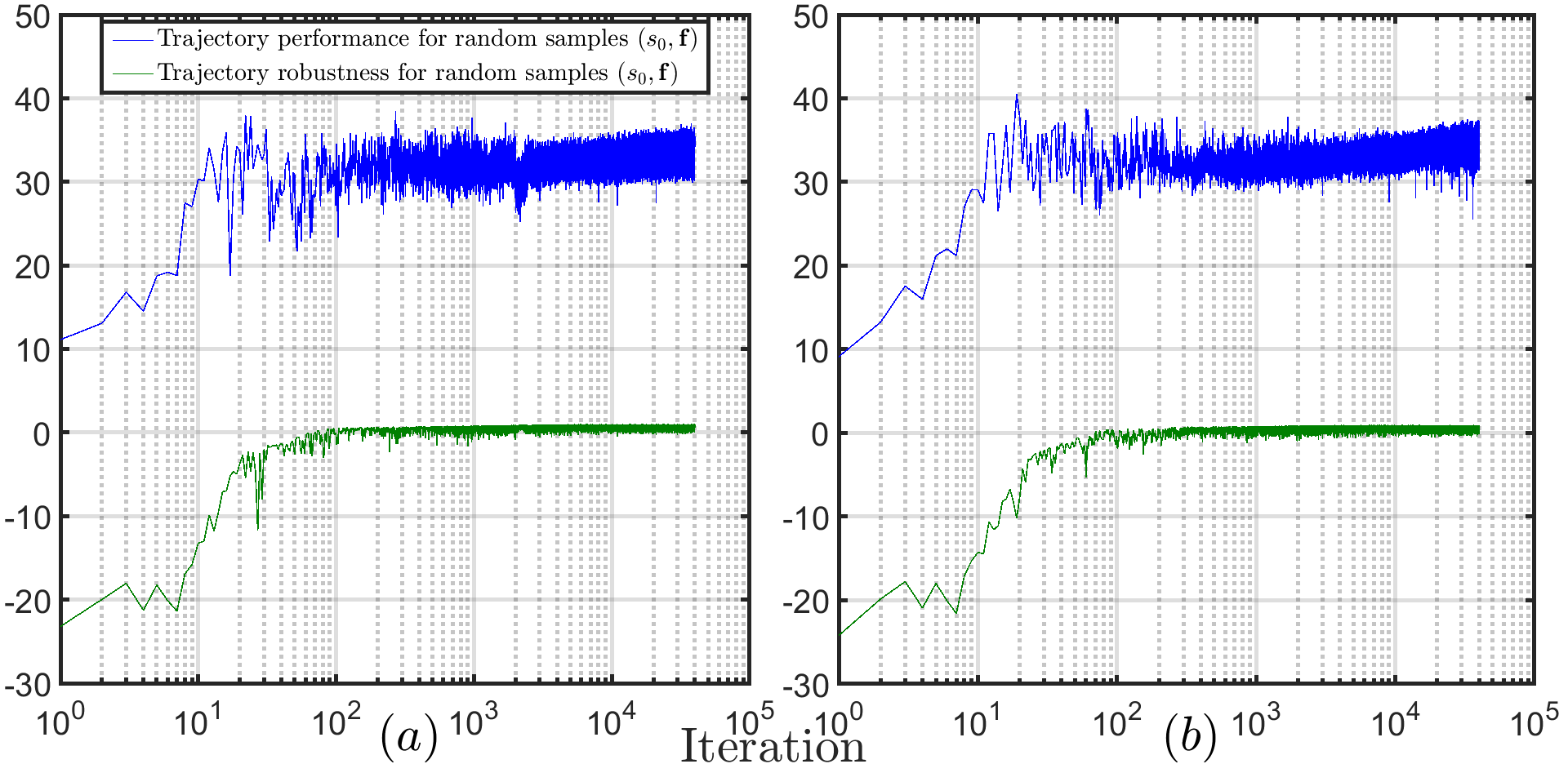}
    \captionof{figure}{\small{Figures (a) and (b) present the evolution of performance cost (blue) vs trajectory robustness (green) over the training process of unicycle dynamics for $\rho=0.5, 0.3$ respectively. The horizontal axis is presented in $\mathbf{log}$ form.}}
    \label{fig:tradeoffs}
    \end{minipage}
    \begin{minipage}{0.44\textwidth}
    \includegraphics[width=0.9\linewidth]{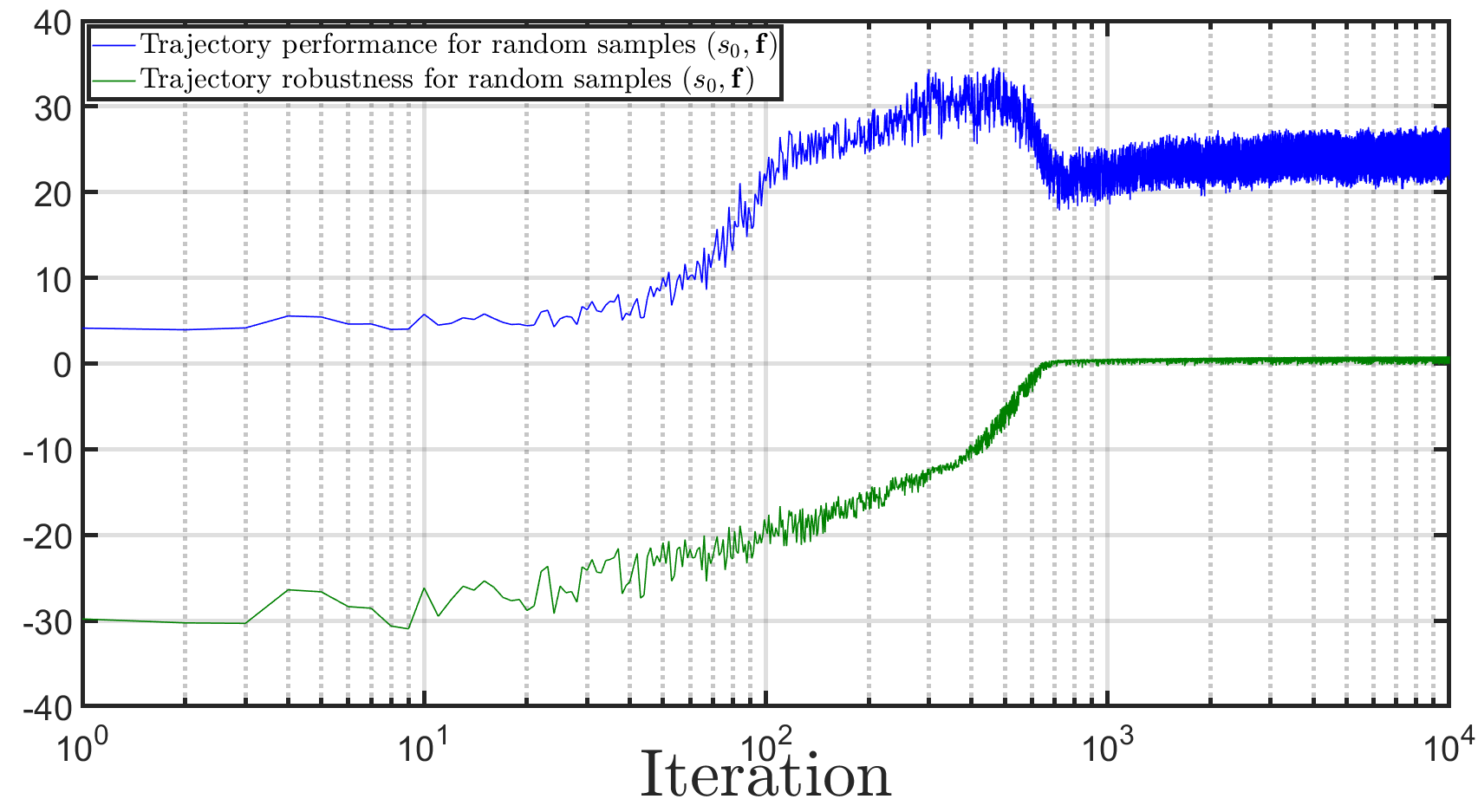}
    \captionof{figure}{\small{Presents the evolution of performance cost (blue) vs trajectory robustness (green) over the training process in quadrotor example for $\rho=0.1$. The horizontal axis is in $\mathbf{log}$ form.}}
    \label{fig:tradeoff_quad}

    \end{minipage}
\end{figure*}

\begin{table*}[t]
\centering
\begin{tabular*}{0.99\textwidth}{@{\extracolsep{\fill}}lllllllll}
\toprule
Example & \muc{6}{Training} & \muc{2}{Validation} \\
\cmidrule{2-7}\cmidrule{8-9}
        &  $\rho$ & $\tau$ & Controller & Activation & Iterations &  Runtime  & Expected value & Expected value \\
        &     &    & Dimension  & Function   &  & (secs)
        & Performance  &  $\STLobjective$ / $\rob_\varphi$ \\
\midrule
Unicycle  & 0.3 & 1e2 & [4,5,2,2] & tanh & 40000 & 1048 & 35.3430 & 0.6108\ / \ 0.6109 \\
Unicycle  & 0.5 & 1e2 &[4,5,2,2] & tanh & 40000 & 1067 & 33.3528 & 0.8456\ / \ 0.8518\\
\midrule
Quadrotor & 0.1 & 5e4 &[7,10,3,3]& tanh & 10000 & 155 & 24.3024 & 0.6729\ / \ 0.7516\\
\bottomrule
\end{tabular*}
\caption{\small{Training and Validation Results}}
\label{tab:results}
\end{table*}

\begin{table*}[t]
\centering
\resizebox{0.99\hsize}{!}{$
\begin{tabular*}{1.5\textwidth}{@{\extracolsep{\fill}}lllllllll}
\toprule
Example & \muc{3}{CBF for atomic propositions,\ $\barrierof{\phi_i}(\state_\timeid, \timeid)$} & {Reward} & {discount} \\
\cmidrule{2-4}
        &  $\phi_1:\ \traj_{\state_0}(\timeid)\in \region_1, \ \timeid \in[1,10]$ & $\phi_2:\ \traj_{\state_0}(\timeid)\in \region_2,\ \timeid \in[1,10]$ & $\phi_3:\ \traj_{\state_0}(\timeid)\notin \region_3, \ \timeid \in[1,20]$ &  & \\
\midrule
Unicycle  & $\resizebox{0.2\hsize}{!}{$1-\frac{2}{3}((x_\timeid-2)^2+(y_\timeid-8)^2)$}$ & $\resizebox{0.2\hsize}{!}{$1-\frac{2}{3}((x_\timeid-8)^2+(y_\timeid-2)^2)$}$ & $\resizebox{0.2\hsize}{!}{$1-\exp(1-\frac{2}{3}((x_\timeid-5)^2+(y_\timeid-5)^2))$}$ & $\resizebox{0.2\hsize}{!}{$10\exp{\left( -\frac{(x_\timeid-8)^2+(y_\timeid-8)^2}{36}\right)}$}$  & 0.9  \\
Quadrotor &$\resizebox{0.3\hsize}{!}{$1-\frac{(x(\timeid)-0.025)^2+(y(\timeid)-0.1)^2+z(\timeid)^2}{0.00023438}$}$ & $\resizebox{0.3\hsize}{!}{$1-\frac{(x(\timeid)-0.1)^2+(y(\timeid)-0.025)^2+z(\timeid)^2}{0.00023438}$}$ & $\resizebox{0.3\hsize}{!}{$1-\exp(1-\frac{(x(\timeid)-0.0625)^2+(y(\timeid)-0.0625)^2)+z(\timeid)^2}{0.00023438})$}$ & $\resizebox{0.3\hsize}{!}{$10\exp{\left( -\frac{(x(\timeid)-0.1)^2+(y(\timeid)-0.1)^2+(z(\timeid)+0.0375)^2}{0.0056}\right)}$}$ & 0.9  \\
\bottomrule
\end{tabular*}
$}
\caption{\small{Shows the CBFs and reward functions we utilize in training process.}}
\label{tab:barriers}
\end{table*}

\begin{table}
\centering
\resizebox{\hsize}{!}{$
\begin{tabular}{@{\extracolsep{\fill}}lllllll}
\toprule
Confidence  & \muc{4}{Unicycle Dynamics} & \muc{2}{Quadrotor Dynamics} \\
\cmidrule{2-5}\cmidrule{6-7}
Threshold        &  \muc{2}{$\rho=0.3$} & \muc{2}{$\rho=0.5$} & \muc{2}{$\rho=0.1$}\\
\cmidrule{2-3}\cmidrule{4-5}\cmidrule{6-7}
$ \confthreshold$       & 
$-\vaar_\confthreshold$ &
$-\cvar_\confthreshold$ &
$-\vaar_\confthreshold$ &
$-\cvar_\confthreshold$ &
$-\vaar_\confthreshold$ &
$-\cvar_\confthreshold$ \\
\midrule
0.95  &  0.246  & 0.132  & 0.540 & 0.417 & 0.527 & 0.455 \\
0.98  &  0.133  & 0.036  & 0.421 & 0.311 & 0.452 & 0.395 \\
0.99  &  0.059  & -0.027 & 0.336 & 0.239 & 0.406 & 0.360 \\
0.999 & -0.133  & -0.191 & 0.121 & 0.067 & 0.305 & 0.284 \\
\bottomrule
\end{tabular}
$}
\caption{Risk measures with one million data points.}
\label{tab:verification}
\end{table}

\begin{figure*}
\begin{minipage}{0.305\textwidth}
\centering
\includegraphics[width=0.85\linewidth]{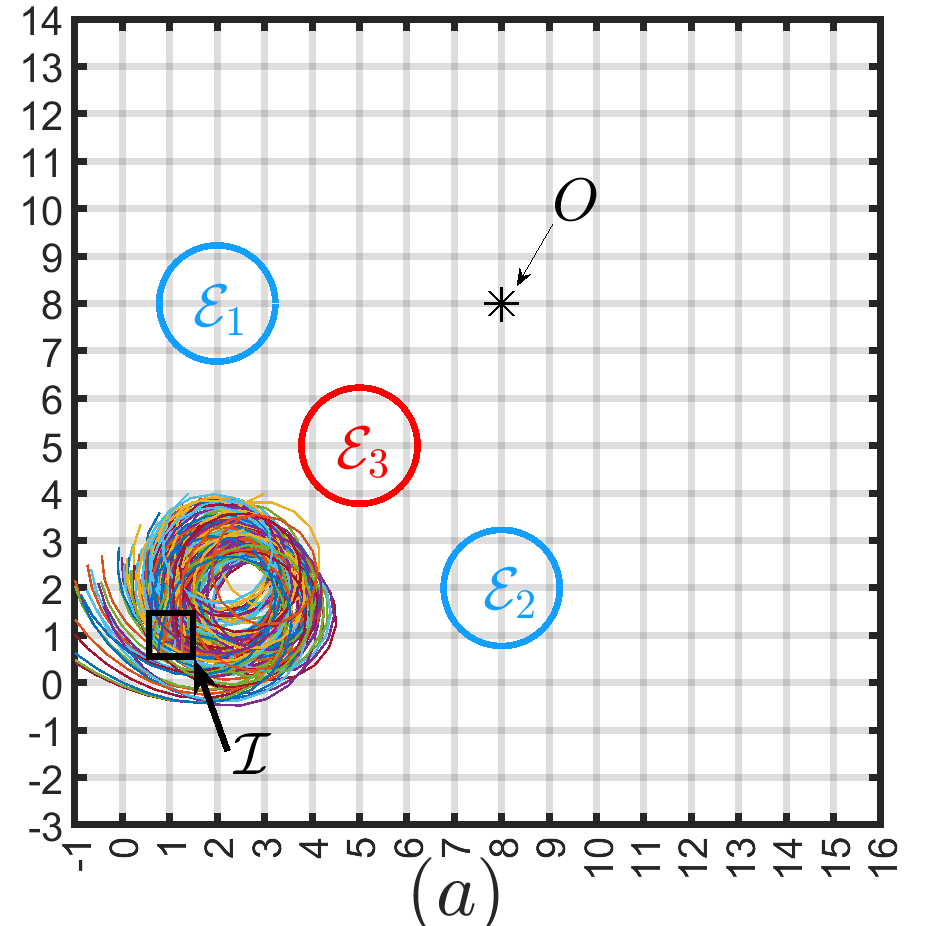}
\end{minipage}
\begin{minipage}{0.312\textwidth}
\centering
\includegraphics[width=0.85\linewidth]{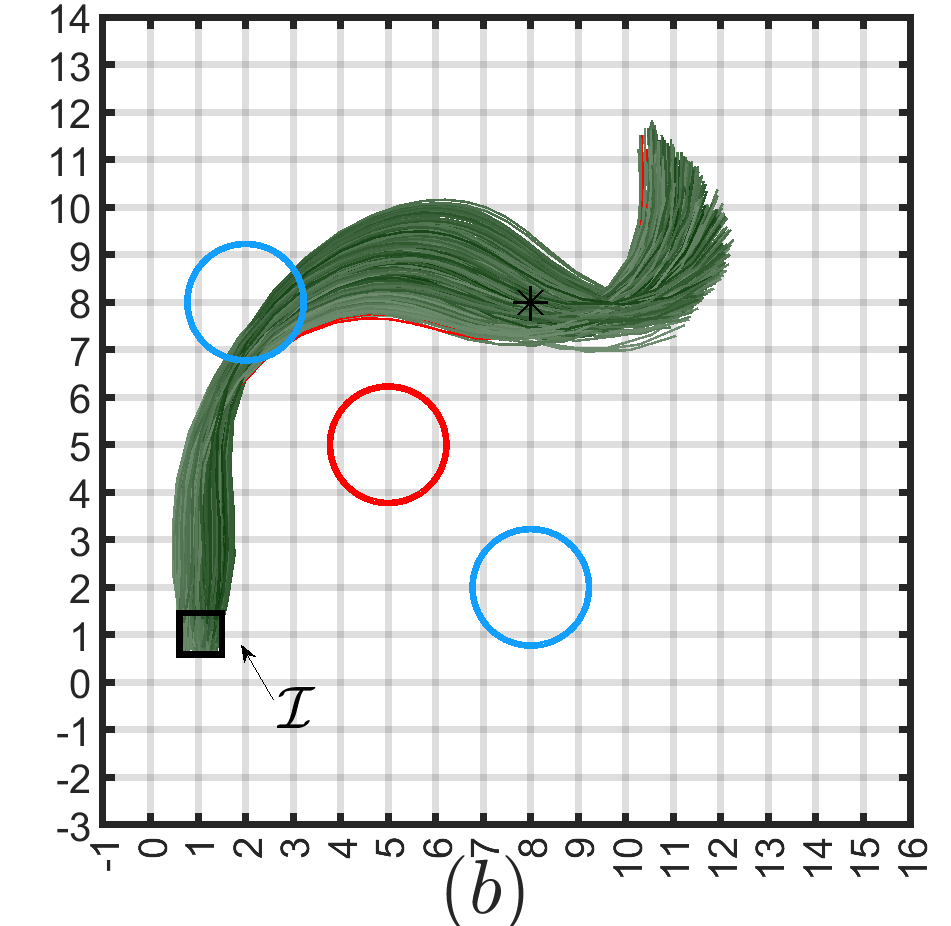}
\end{minipage}
\begin{minipage}{0.374\textwidth}
\centering
\includegraphics[width=0.87\linewidth]{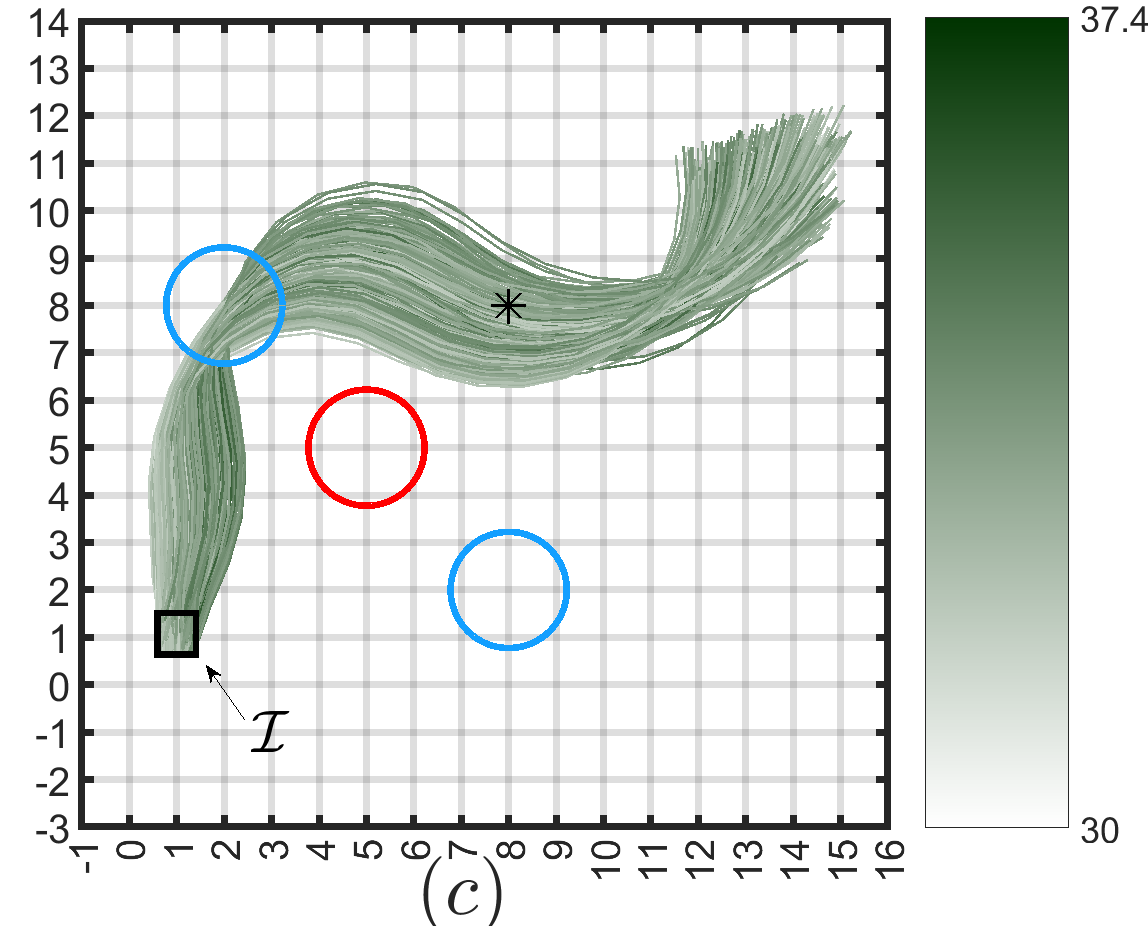}
\end{minipage}
\caption{\small{(a) Represents sample trajectories with the random initial value for $\param$, (b,c) respectively show sample trajectories for trained $\param$ with robustness margin $\rho=0.3$ and $0.5$. This figure clearly shows the trajectories shift towards the center of $\region_1$ when the robustness margin $\rho$ increases. For this simulation we sample 500 different $(\state_0, \dynamics)$ uniformly at random from $(\init\times \modelfamily)$ and simulate the trajectories. The green plots satisfy the STL specifications while its darkness shows the level of performance. There exists $2$ red trajectories in (b) that are marginally violating the STL specs.}}
\label{fig:startvsend}
\end{figure*}

\subsection{Quadrotor Dynamics}
In another attempt we consider controlling a quadrotor with uncertain dynamics. We define the uncertainty for the initial condition as a spherical set, $\init=\mathcal{B}_r(c)$ with center, $c=[0.025, 0.025, 0, 0, 0, 0]^\top$ and radius $r=0.0125$. The quadrotor also follows the following uncertain dynamics,
$$
\label{eq:QR}
    \begin{bmatrix}x(\timeid+1)\\ y(\timeid+1)\\ z(\timeid+1)\\ v_x(\timeid+1)\\ v_y(\timeid+1) \\ v_z(\timeid+1)\end{bmatrix}=\begin{bmatrix}(1+\delta)x(\timeid)+0.05v_x(\timeid)\\(1+\delta)y(\timeid)+0.05v_y(\timeid)\\(1+\delta)z(\timeid)+0.05v_z(k)\\ (1+\delta)v_x(\timeid)+0.4905 \tan(u_1(\timeid))\\(1+\delta)v_y(\timeid)-0.4905 \tan(u_2(\timeid)) \\ (1+\delta)v_z(\timeid)+0.05(\mathrm{g}-u_3(\timeid))  \end{bmatrix},
$$
\noindent discretized with ZOH for timestep $T=0.05$ sec. Here $\delta \in \left[-0.01,0.01 \right]$ and the control inputs, $u_1(\timeid) \in [-0.1,0.1],\ u_2(\timeid) \in [-0.1,0.1], \ u_3(\timeid) \in [7.81,11.81]$. The parameter $\mathrm{g}=9.81$ is the gravity. To impose bounds on the controller, like the Unicycle example, we fix the last hidden layer of the neural controller, $[\tanh,\tanh,\tanh]$ and include it in the model,
$$
\begin{aligned}
 &u_1(\timeid) &\gets\   &0.1 \tanh(0.1 a_1(k)), \ &a_1(k)\in \mathbb{R}\\
 &u_2(\timeid) &\gets\   &0.1 \tanh(0.1 a_2(k)), \ &a_2(k)\in \mathbb{R}\\
 &\mathrm{g}-u_3(\timeid) &\gets\   &2 \tanh(0.1 a_3(k)), \ & a_3(k)\in \mathbb{R}
\end{aligned}
$$


\subsection{Results}
\begin{figure*}
\centering
\includegraphics[width=0.85\linewidth]{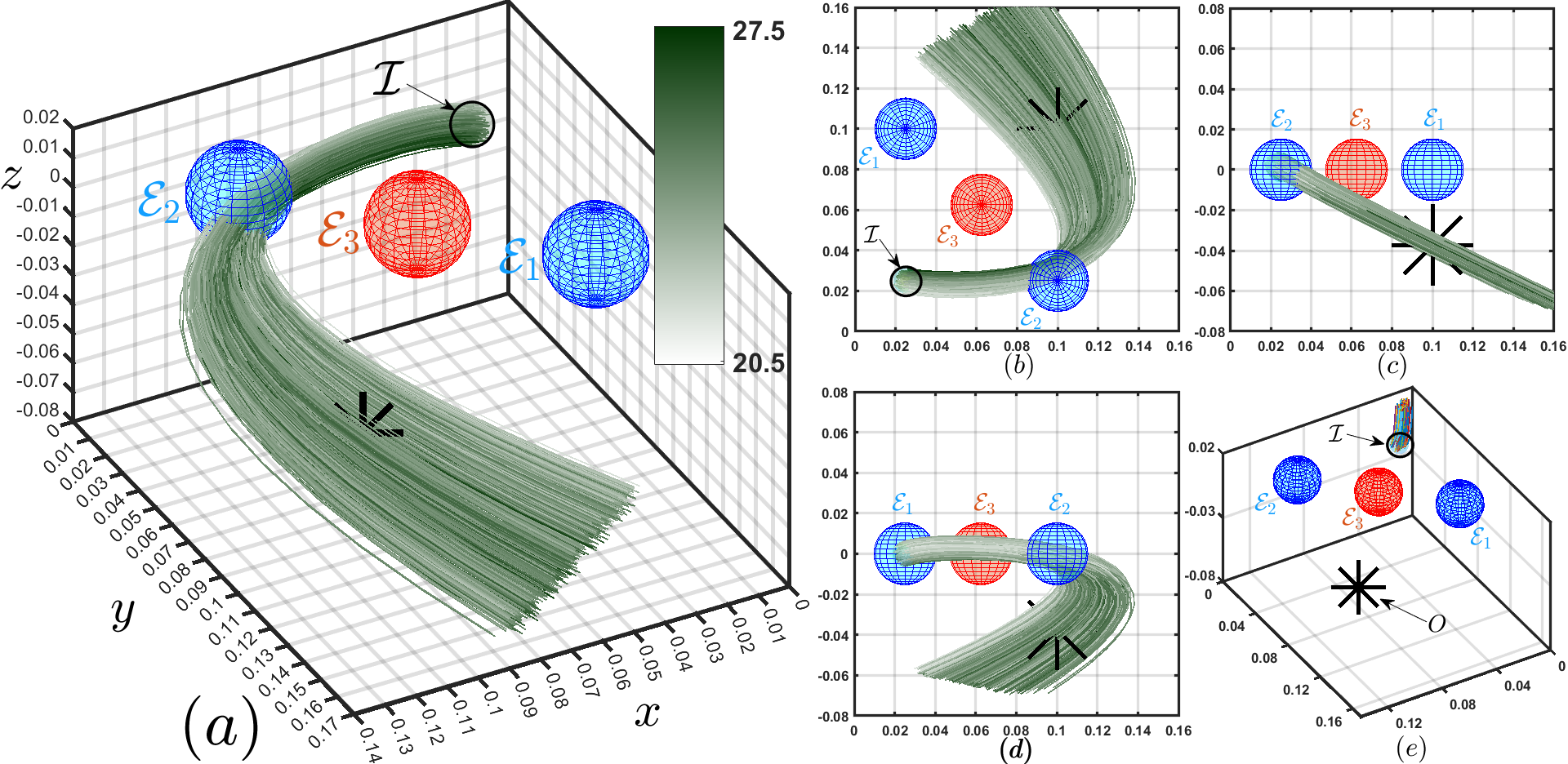}

\caption{\small{(a) Represents 500 trajectories generated with trained controller parameters $\param$ for $\rho=0.1$. For this simulation, we sample 500 different $(\state_0, \dynamics)$ uniformly at random from $(\init\times \modelfamily)$ and simulate the trajectories. The darkness of trajectories is corresponding to their level of performance. There is no trajectory violating the STL specification. (b,c,d) shows the projection of trajectories on {\bf X-Y, Y-Z} and {\bf X-Z} planes respectively. (e) Represents simulated sampled trajectories for the initial value of $\param$ that we utilized in training process.}}
\label{fig:startvsend_quad}
\end{figure*}

The STL specifications for both examples are adopted from
\cite{liu2021recurrent} and are introduced in Eq.~\eqref{eq:stlspec}
(Example~\ref{ex:one}). Regions $\region_1,\region_2$ and $\region_3$
for unicycle and quadrotor examples are introduced in
Fig~\ref{fig:startvsend} and Fig.~\ref{fig:startvsend_quad}
respectively. The unicycle and quadrotor approaches to the target
$O=\transpose{[8 , \ 8]},\ O=\transpose{[0.1 , \ 0.1, \ -0.0375]}$
respectively. They are planned to approach $O$ with the highest
possible level of performance (fast and close) within $\horizon=20$
time steps. The reward function and CBFs are defined in
table~\ref{tab:barriers} for both examples. We train a controller to
satisfy the performance and STL task for unicycle and quadrotor
dynamics. Table~\ref{tab:results} shows the training result for
$\rho=0.3,0.5$ in unicycle and $\rho=0.1$ in quadrotor example. This
table shows the trade-off between performance and STL robustness for
the unicycle example. 

\begin{figure*}
    \begin{subfigure}{0.49\textwidth}
    \centering
    \includegraphics[width=0.75\linewidth]{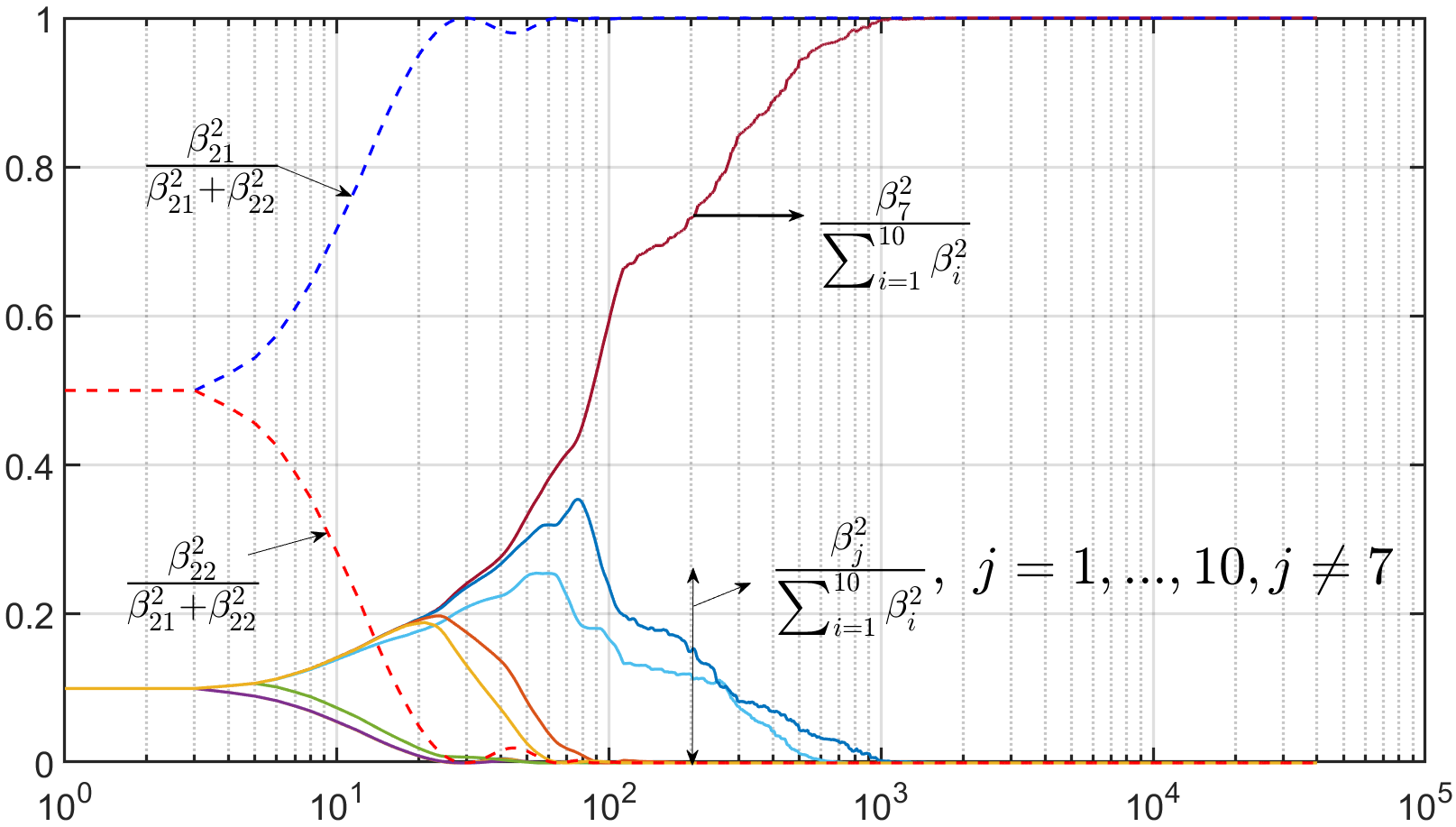}
  \caption{}
    \end{subfigure}
    \begin{subfigure}{0.49\textwidth}
    \centering
  \includegraphics[width=0.75\linewidth]{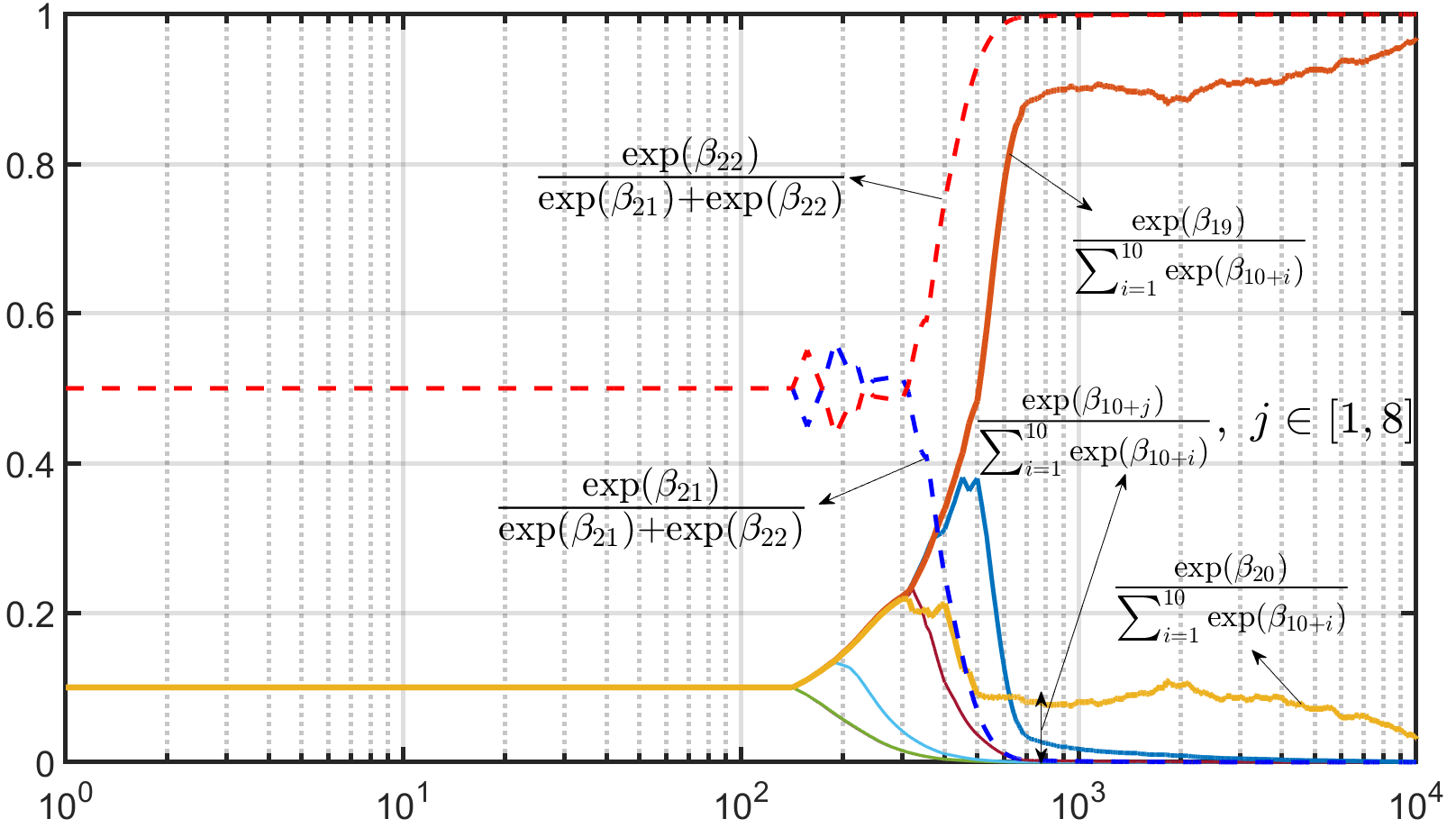}
  \caption{}
\end{subfigure}
\caption{\small{(a,b) shows the evolution of disjunctive parameters over 
the training process in unicycle ($\rho=0.3$) and quadrotor ($\rho=0.1$)
examples, respectively. The log-scale horizontal axis indicates number 
of training iterations. There are three disjunctive formulas in
\eqref{eq:stlspec}: $\ev_{[1,10]}(\state \in \region_1)$,
(that needs parameters $\beta_1,\ldots,\beta_{10}$) in its CBF, 
$\ev_{[1,10]}(\state \in \region_2)$ (using parameters $\beta_{11},\ldots,
\beta_{20}$ and the disjunction between these formulas that uses parameters 
$\beta_{21}$ and $\beta_{22}$. %
(a) Parameter $\beta_{22}$ converging to zero indicates that the system 
chooses to satisfy the first subformula thus the variables $\beta_{11}, 
\ldots, \beta_{20}$ are not relevant and not plotted. The $\beta_7$ parameter has the largest value, indicating the majority of the trajectories are in 
region $\region_1$ at time $\timeid = 7$.%
(b) Here, $\exp(\beta_{21})$ converging to zero implies that $\beta_{1},\ldots,\beta_{10}$
are not relevant. As the parameters $\beta_{19}, \beta_{20}$ are nonzero,
the majority of trajectories are in $\region_2$ at $\timeid=9$ 
and the others at $\timeid=10$.}}
 
    \label{fig:beta_value}
\end{figure*}

We utilized $\eqref{eq:wtavg1}$ and \eqref{eq:wtavg2} in
training the disjunctive parameters $\bbeta$ for unicycle and
quadrotor respectively. Fig.~\ref{fig:beta_value} shows the evolution
of disjunctive parameters over the training process.
Fig.~\ref{fig:tradeoffs} and \ref{fig:tradeoff_quad} present the
trade-off between the performance cost $\perfobjective$ and trajectory
robustness $\STLobjective$ over the training process for both
examples. Fig.~\ref{fig:startvsend} and Fig.~\ref{fig:startvsend_quad}
present the simulation of trajectories for unicycle and quadrotor
examples, respectively. 

Table~ \ref{tab:verification} presents the results on probabilistic
verification or risk-analysis for the controllers. For the unicycle
dynamics, we can see that increasing the robustness margin parameter
$\rho^*$ leads to
an increase in the (probabilistic) lower bound on the robustness.
Increasing the confidence level reduces the probabilistic lower bound.
In fact, at 99.9\% confidence, there is a risk of seeing system
behaviors that violate the specifications by a margin of $0.133$.
Similar risks can be seen at the 99\% and 99.9\% confidence in the
$\cvar$ values. Intuitively, table~\ref{tab:verification} matches our
expectation that controllers designed with higher robustness margin
should have lower risk of violating specifications (at the cost of
performance).

\ignore{
\begin{figure*}
    \begin{subfigure}{0.49\textwidth}
    \centering
    \includegraphics[width=\linewidth]{figures/reward_dist_stack.eps}
  \caption{}
    \end{subfigure}
    \begin{subfigure}{0.49\textwidth}
    \centering
      \includegraphics[width=\linewidth]{figures/rob_hard_dist_stack.eps}
    \caption{}
    \end{subfigure}
    \caption{\small{(a) Makes a comparison for the distribution of performance obtained from training process between $\rho=0.3,\rho=0.5$ in unicycle example. (b) Presents a comparison for the distribution of trajectory robustness between $\rho=0.3,\rho=0.5$ in unicycle example. }}
    \label{fig:dist}
\end{figure*}
}

\section{Conclusion and Future work}
\label{sec:relwork_conc}

\noindent In this work we propose the weighted average, a useful tool to include disjunctive STL formula in the existent soft constrained policy optimization techniques \cite{lindemann2018control}. We also utilize time dependent feedback policies that facilitates control in presence of STL specifications. This enables us to control the model with smaller neural networks. Non-convex optimizations may be intractable for Lagrange multiplier techniques. We address this problem with proposition of a training algorithm that simulates the trade off between objective and its constraints. We finally utilize this training algorithm for non-convex policy optimization with respect to STL specifications.

\noindent In the future, we will focus on improving the scalability of the training process. The proposed recurrent structure for feedback models suffers from vanishing or exploding gradient issue. This results in inefficient training for long trajectories and is due to its resemblance to RNN structures. Thus we plan to include LSTM structure with introduction of hidden states between feedback blocks in the recurrent dynamic structure.

\bibliographystyle{IEEEtran}
\bibliography{main}
\end{document}